# The Digital Public Library of America Ingestion Ecosystem: Lessons Learned After One Year of Large-Scale Collaborative Metadata Aggregation

Mark A. Matienzo
Digital Public Library of America, USA
mark@dp.la

Amy Rudersdorf
Digital Public Library of America, USA
amy@dp.la

## Abstract

The Digital Public Library of America (DPLA) aggregates metadata for cultural heritage materials from 20 direct partners, or Hubs, across the United States. While the initial build-out of the DPLA's infrastructure used a lightweight ingestion system that was ultimately pushed into production, a year's experience has allowed DPLA and its partners to identify limitations to that system, the quality and scalability of metadata remediation and enhancement possible, and areas for collaboration and leadership across the partnership. Although improved infrastructure is needed to support aggregation at this scale and complexity, ultimately DPLA needs to balance responsibilities across the partnership and establish a strong community that shares ownership of the aggregation process.

## 1. Introduction

The Digital Public Library of America (DPLA) recently celebrated its first anniversary aggregating the riches of America's libraries, archives, and museums and sharing them through a single portal. With its Hubs (the 20 direct partners from whom DPLA harvests records) and their partners (approximately 1,300 in all), DPLA has worked to make these resources freely available to the world. After a year focusing resources on growth, with the DPLA holdings more than tripling to over seven million records in twelve months, it seems an appropriate time to take stock of the technologies and processes within which this work occurs, as well as the data models used to aggregate the Hubs' various metadata standards and the nature of collaboration between DPLA and the Hubs. It is important to identify areas both of success and improvement that have become apparent since the launch in April 2013. This assessment takes into consideration outside variables, as well, including feedback from Hubs, users of DPLA's open and freely available application programming interface (API), and others interested in the DPLA technology stack and metadata model. A few areas of future work have been identified, which will help to create a roadmap for ongoing investigation and development. It is hoped, too, that this process will involve current and future partners, and create a community of practice around these open source technologies and metadata management systems.

## 2. Development, implementation, and current status of DPLA infrastructure

DPLA launched its services on April 18, 2013, with 2.4 million records from 16 Hubs (and their over 900 partners) after a two-year planning phase. The components that make up the technology stack that supports the infrastructure are lightweight and open source, which allowed DPLA's initial technical implementation team to prototype and deploy working iterations quickly. DPLA also developed a metadata application profile, or MAP (Digital Public Library of America, 2014a), based on existing data standards and models. In addition to the ingestion system described below, DPLA's infrastructure also provides both an application programming interface (API) and a public user interface that serves as the primary discovery system for the ingested metadata. The platform, or API layer, is a Ruby on Rails web application that provides an abstraction mechanism over the primary data store and search index. The portal, or user-facing front-end application, is built on Ruby on Rails, and is a client of the platform application.

The DPLA technical infrastructure was implemented over a period of 18 months, which demanded a relatively short build-out process. During the initial implementation period (October 2012-April 2013), the DPLA Assistant Director for Content undertook primary responsibility for developing the metadata mappings, and a team of contractors developed the metadata ingestion system and other areas of infrastructure and ran the ingestion processes. Since late 2013, the DPLA staff has steadily grown, including the hiring of a Director of Technology (December 2013), two Technology Specialists (January and May 2014), a Data Services Coordinator (August 2014), and a Metadata and Platform Architect (August 2014). During this time, DPLA has undertaken most of the responsibility for maintaining the existing infrastructure, overseeing the ingestion process, and identifying areas for improvement.

## 2.1 The DPLA Metadata Application Profile

The DPLA Metadata Application Profile (MAP) is an extension of the Europeana Data Model, or EDM (Europeana, 2014). Version 3, the first public version of the MAP, was developed in early 2013 by DPLA staff and others, in collaboration with Europeana staff and public data specialists who provided input during an open review period in late 2012. Like EDM, the MAP incorporates or references a variety of standards and models, including the Dublin Core Metadata Element Set, Dublin Core Terms, the DCMI Type Vocabulary, OAI-Object Reuse and Exchange, and others. While based on EDM, the DPLA MAP nonetheless slightly diverges from it. First, one of the MAP's core classes, the Source Resource (dpla:SourceResource), is defined as a subclass of the corresponding class in EDM (Provided Cultural Heritage Object, or edm:ProvidedCHO). The primary motivation for this was to make clear that the properties of dpla:SourceResource in some cases may have different cardinalities or requirements than those defined for edm:ProvidedCHO. In addition, because of limitations on both the data available from DPLA's providers and the geocoding enrichments implemented near launch, DPLA developed its own spatial location class, dpla:Place.

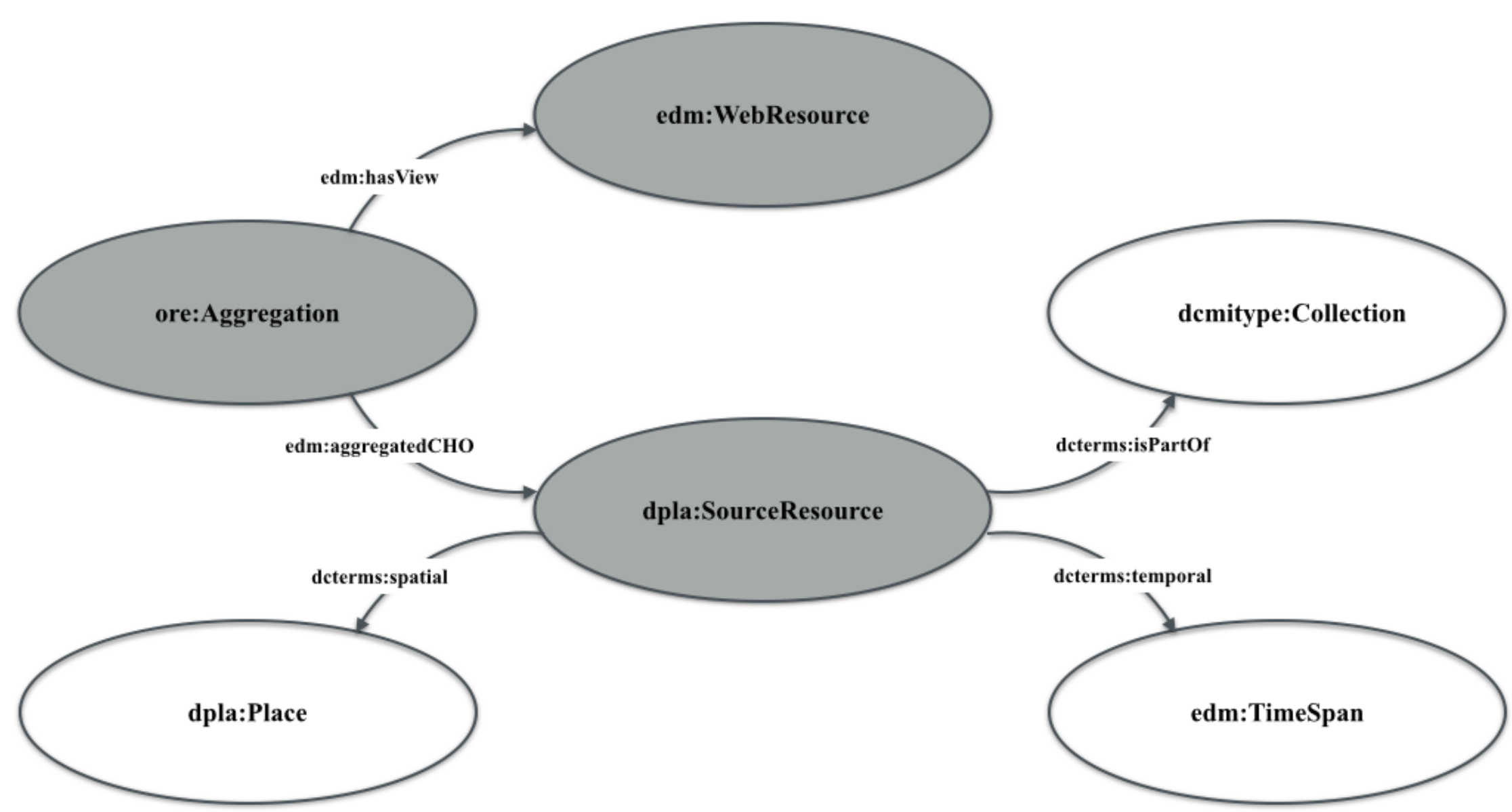

FIG. 1. Core classes and relationships in the DPLA Metadata Application Profile, versions 3 and 3.1.

DPLA staff reviewed and revised the requirements for the MAP in mid-2014, and released MAP version 3.1 in July 2014. Many of the differences between MAP versions 3 and 3.1 relate to cardinality requirements, which were changed based on recognition of the properties DPLA could not reliably receive, map, or otherwise derive from metadata provided by Hubs. DPLA also added a new property (Intermediate Provider, or dpla:intermediateProvider) to allow for the declaration of an entity understood to be distinct from the two provider-related properties within EDM (edm:Provider and edm:dataProvider). MAP version 3.1 defines an Intermediate Provider as "an intermediate organization that selects, collates, or curates data from [an edm:dataProvider] that is then aggregated by [an edm:Provider] from which DPLA harvests" (Digital Public Library of America, 2014a). Beyond these changes, MAP version 3.1 also contains several changes which bring it towards further alignment with EDM, such as clearly identifying the super-properties for a given property when available, aligning internal properties with EDM definitions, adding the edm:hasType property to express genre statements, and adding the edm:rights property. The addition of edm:rights allows for association of rights information available at from a given URI to two core classes within the MAP.

## 2.2 Ingestion system and workflow

The DPLA ingestion system (Digital Public Library of America, 2014b) is an application, written in Python using the Akara (2010) framework, that provides REST endpoints for web services to transform or enrich data serialized in JSON. The primary DPLA data store is a BigCouch/CouchDB document-oriented database, with metadata both stored and serialized using JSON-LD 1.0 (Sporny, Kellogg, and Lanthaler, 2014). Once stored in BigCouch, all ingested metadata is indexed using Elasticsearch, a REST-based search server built upon Apache Lucene. Additional scripts that support or control the ingestion process are also written in Python. The ingestion workflow for a given *ingestion source* has a designated *ingestion profile*. In most cases, Hubs only provide one ingestion source, but a small number of Hubs are continuing to develop internal systems to support the single-ingestion-source model that is, technically, a requirement to DPLA participation. Accordingly, a single Hub that has more than one ingestion source may have multiple ingestion profiles. Each ingestion profile is a JSON document containing configuration information such as the type of harvest, (e.g., OAI-PMH, site-specific API, static files, etc.), location of an HTTP endpoint if applicable (e.g., the OAI-PMH provider URI), the specific mapping and enrichments to be applied, and other internal settings required by the ingestion system.

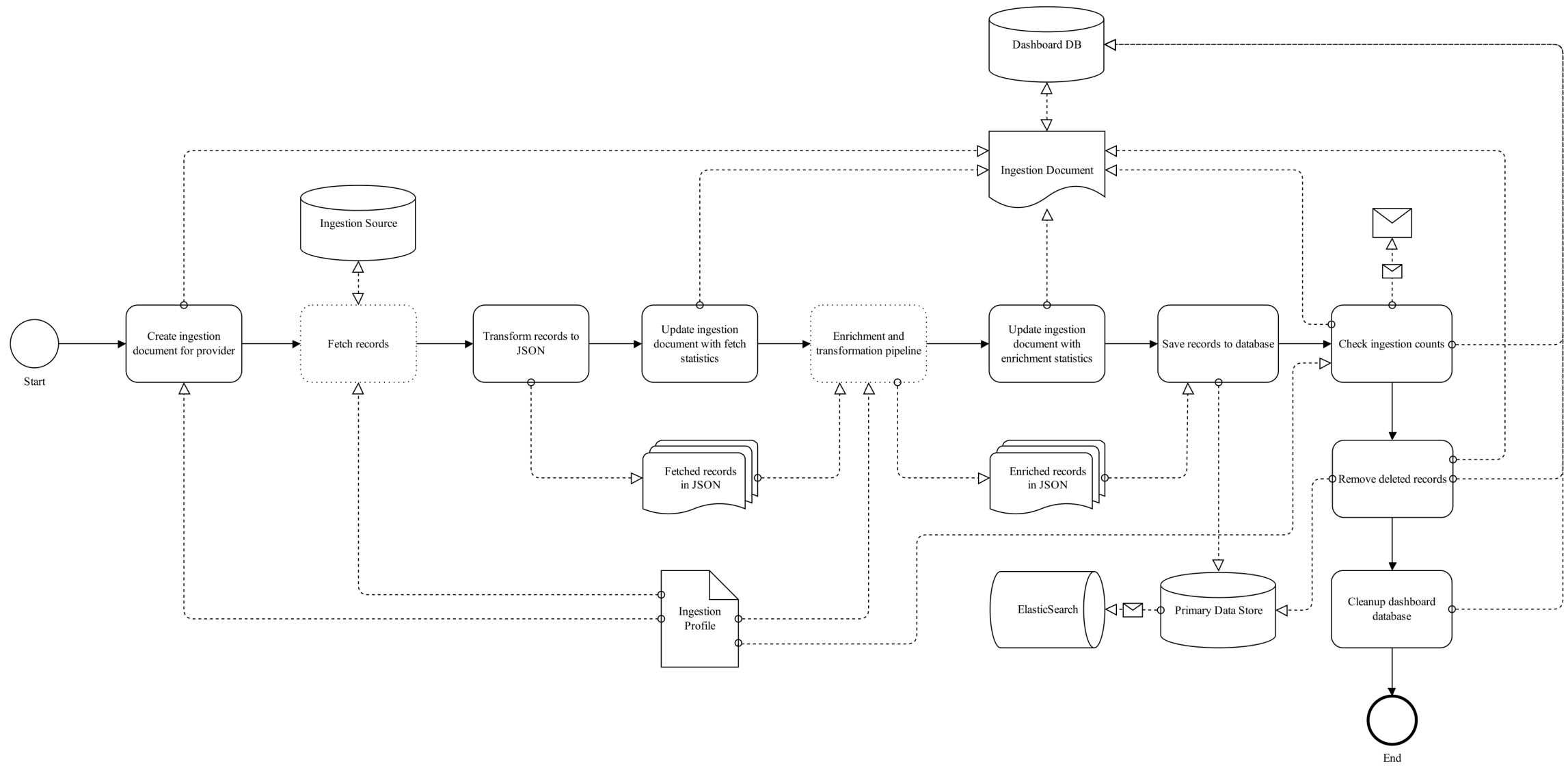

FIG 2. Overview of the DPLA ingestion workflow.

The ingestion workflow is invoked by a support script that reads the ingestion profile for a given source and creates an *ingestion document* in the *dashboard database* for a given *ingestion process*. The ingestion document contains data about the state of particular *ingestion task* (e.g., whether a specific step has started, completed, or failed). The dashboard database also temporarily contains a representation of each fetched record to allow staff to identify what parts of an ingested record have changed. Once the ingestion document is created, the staff running the ingestion process invokes the *fetch task*, which obtains the metadata to be ingested from the source defined in the profile. The metadata is then deserialized from its native format (typically XML), reserialized as a JSON expression of the original data, and persisted to disk in a temporary location. Once the fetch process is complete, the ingestion document is updated to contain the location of the data transformed to JSON.

The ingestion staff then invokes the *transformation and enrichment tasks*. These tasks map and transform the JSON-serialized metadata to the DPLA MAP, and normalize, enhance, and augment the metadata using a "pipeline" that orchestrates requests to the application's REST endpoints (see section 2.3 for more information). Once complete, the records are temporarily persisted to disk as a JSON-LD serialization of the MAP, and the ingestion document is updated with information about transformation and enrichment processes, including location of the transformed records and the extent of any failures within the process. The ingestion staff then runs the *save task*, which reads the MAP-compliant JSON-LD records and persists them to the primary data store. After the save process completes, the ingestion staff runs the *check ingestion counts task*, which identifies the number of new, updated, or deleted records for each ingestion process and automatically alerts the identified staff when those values are above a certain threshold defined in the ingestion profile. Finally, the ingestion staff runs two concluding tasks: the *remove deleted records task* and the *dashboard database cleanup task*. Both tasks remove objects from the primary data store or dashboard database. These objects correspond to the metadata from ingested records that were either deleted from the ingestion source by the provider (e.g., as identifiable using the <deleted> element from an OAI-PMH provider) or otherwise not present or available during a given ingestion process.

## 2.3 The metadata transformation and enrichment pipeline

Most of the work to transform, normalize, and enhance the metadata ingested into DPLA occurs as part of the *transformation and enrichment pipeline*, which executes a list of specific steps defined in an ingestion profile in a specific, linear order. Each of the steps is implemented in the ingestion system as a module mounted at a defined REST endpoint. Each of the endpoints receives JSON data over an HTTP POST request, and returns JSON data, either modified if the step was applicable and successful or unchanged if the step was inapplicable or if it failed. Most of the ingestion profiles share a number of common steps, and the modular design allows DPLA to easily reuse them and add extra parameters as needed.

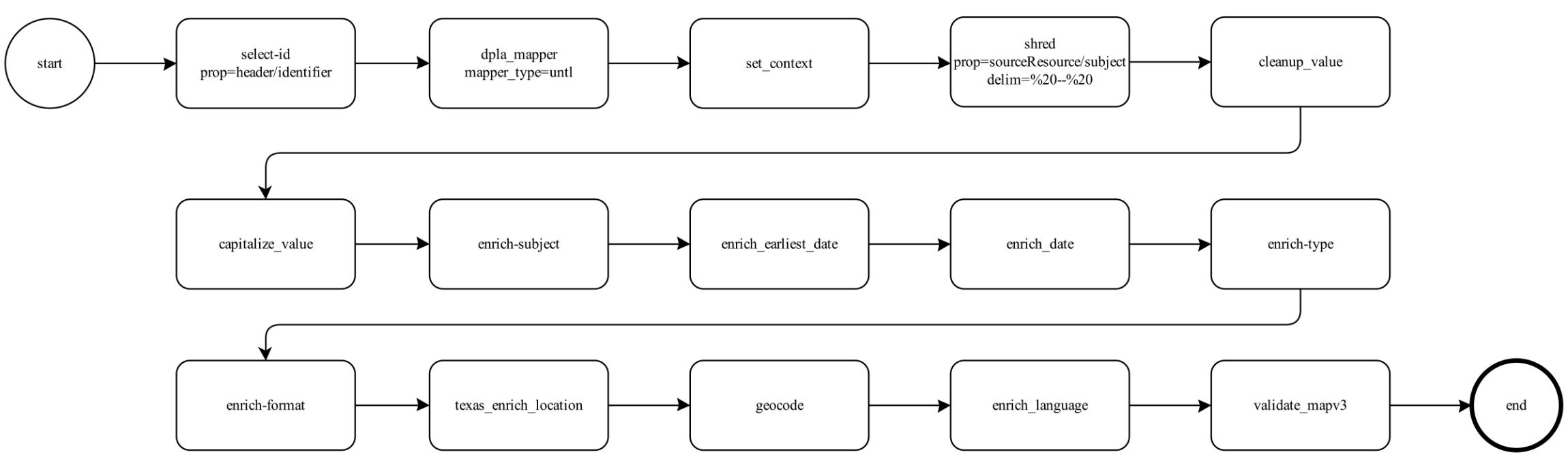

FIG 3. Sample transformation and enrichment pipeline for ingestion from the Portal to Texas History.

At a minimum, the pipeline must contain two steps: one that selects the source of the identifier from the ingested metadata (which is required for persistence), and another that transforms and maps the metadata to the DPLA MAP. Despite the pressures related to launch, DPLA was also able to implement some degree of normalization and enrichment. Much of the DPLA staff's ongoing work involves revising and ensuring that these normalization and enrichment modules remain robust and error-free. At a minimum, the enhancements applied to most metadata ingested into DPLA include what Hillmann, Dushay, and Phipps (2004) term "safe transforms," through global cleanup of values to address minor differences in capitalization, punctuation, or whitespace, or alignment and reconciliation of terms against comparatively small controlled vocabularies such as the DCMI Type Vocabulary or ISO 639-3 language codes. In addition, the ingestion system undertakes more complex transformations based on diversity of practice, such as normalizing dates or date ranges to a common format, and "shredding" a string literal based on a given delimiter to yield multiple values. In addition, the ingestion system also includes a geocoding enrichment service, which uses external services to take geographic name values and geocode them to return latitude and longitude pairs, and then uses those coordinates to build out a geographic hierarchy. More details about these services are provided below.

The quick lead up to the launch meant turnaround times were limited and the need to ingest metadata created using different schemas under varying practice and assumptions meant that some areas of work on the transformation and enrichment pipeline had to be reprioritized. Work during the initial ingest, which took place roughly between February and mid-April 2013, focused on mapping and the conceptual alignment of fields from the initial 16 Hubs, rather than on the review and quality control of the actual values. Likewise, a loosening of validation against the MAP assertions was necessary to ensure that goals and timelines were met. This period focused on return on investment in the strictest sense: providing the best data in the shortest period of time with the least remediation. In addition, since MAP version 3 was only finalized approximately three months before launch (and only days before the first ingests began), additional changes to the ingestion code and DPLA's Platform API were necessary to ensure that all of the data was available through the portal by mid-April 2013.

## 3. Concerns and challenges

The technology and data model established for the launch has served DPLA well. It has effectively aggregated over seven million records, enabling hundreds of users to utilize the API and effectively build apps, and more than a million users to search and enjoy the resources available through the portal. With sustained use and the ongoing need to continue the ingestion of metadata from both current and future Hubs, challenges have arisen that signal a need to consider potential new options for aggregation, storage, and delivery.

### 3.1. The ingestion process

Ingest remains a very hands-on endeavor. Once a Hub's data is mapped to the DPLA Metadata Application Profile (by the Assistant Director for Content, at the time of publication), a new ingestion profile is written (by DPLA technology staff) that delineates the harvesting, transformation and enrichment steps. In addition, despite using common metadata standards (e.g., DCMES or MODS) or harvesting protocols (e.g., OAI-PMH), differences in local implementation often require DPLA technology staff to modify or supplement implemented mappings, employ new transformation services, or resolve other inconsistencies before an ingest moves to production. For example, several Hubs have found it difficult to reliably provide URIs for thumbnail images for the items associated with the metadata ingested by DPLA. As this information is mandatory in MAP version 3.1, DPLA technology staff must often undertake a degree of reverse engineering to add an enrichment step that identifies or constructs this URI. Nonetheless, while discussions between Hub and DPLA personnel lead to good results, the process of getting a new data set into production often lasts between four and eight weeks.

The ingestion process itself is also resource intensive, and as described above, the architectural paradigm of the current ingestion system currently expects that a consistent transformation and enrichment pipeline be used across *all* ingestion processes from a given ingestion source. A large number of processes are applied to all incoming ingests regardless of the metadata schema used or quality of the metadata received. Currently, data from each Hub is reingested *in its entirety* monthly, every other month, or quarterly, depending on the frequency of local updates. Accordingly, each step defined in the transformation and enrichment pipeline runs during each ingestion process. This ultimately leads to the potential for some enhancements to be lost or misapplied if a Hub has modified its metadata in the interim. Improved control over the enrichment workflow, such as enabling or disabling certain processes for a scheduled ingestion process for a specific Hub, and supplementing those enrichments with provenance information, could provide better control and reduce complexity of ingestion on an ongoing basis. And while the process has been internally standardized, it remains somewhat opaque to some Hubs, especially those who may not be familiar with the languages in which the transformation and enrichment pipeline modules are written. In the experience of DPLA, this also points to the need for improved unit tests and documentation that make the intent of the pipeline modules clearer to domain experts without programming knowledge.

Other challenges to the current model that have come to light over the past year include the inconsistency of some of the enrichment and normalization processes that are applied to all collections. For example, DPLA staff recently identified that structured spatial information (i.e., a place hierarchy) provided by some Hubs was not successfully mapped to the property required for the literals to appear in the user interface (skos:prefLabel). Diagnosis of issues in the enrichment process proves to be an ongoing challenge for DPLA given that the ingestion system does not track the provenance of statements created or modified during transformation and enrichment. In addition, while the DPLA MAP is a data model based upon RDF, the current infrastructure has not yet implemented a complete expression of the constraints defined by it. These limitations originate mostly because the current implementation of validation relies on a simplified expression of the MAP using JSON Schema (Galiegue, Zyp, and Court, 2013), with any validation of the statements about a given item against the MAP currently limited to cardinality checks and simple controlled value verification based on the JSON serialization of the data.

Another area in which DPLA continues to face challenges is the geocoding enrichment process, which retrieves a "best guess" set of coordinates for a term from the Bing Maps API, and uses those coordinates to build out the rest of a geographic hierarchy for that term using the Geonames API. For the value "Charlotte (NC)," the values "35.226944, -80.843333" are automatically assigned via the Bing Maps API to indicate the geographic center of the city. Then, those coordinates are sent to the Geonames API to extract the geographic hierarchy for Charlotte, i.e., United States -- North Carolina -- Mecklenberg County -- Charlotte. This is rich and valuable data that allows DPLA to plot "Charlotte (NC)" on the interactive map in the portal. Like any scaled transformation, this process is not fail-safe, as a careful study of the map exposes. For example, consider a record with the spatial value of "Wisconsin." In this model, the coordinates for the central point of the state identify a hierarchy that contains county-level information (United States -- Wisconsin -- Portage County), which introduces data that can be misleading, if not erroneous. In addition, DPLA staff has discovered that external web services like the Bing Maps API often update the data they provide or their indexing mechanism, which has led to inconsistencies in the geocoding enrichment processes over time. Considering the lack of confidence about the geocoding process and the inability to track provenance of statements in DPLA's current infrastructure, DPLA has chosen not to implement reconciliation of geographic names with URIs from sources such as Geonames until these issues can be addressed.

### 3.2. The metadata

Over the past year, DPLA staff has had the opportunity to work closely with Hubs from across the United States. Not surprisingly, the Hubs employ various metadata standards, maintain data in many different repository types, and manage localized workflow models. The process of aggregation, and especially enrichment and normalization, has been eye-opening for most of the parties involved. DPLA staff knew even before harvesting began in early 2013 that the process would be complex and not without challenges, as evidenced by past work on projects such as the National Science Digital Library (Lagoze et al., 2006), the Digital Library Federation Aquifer Project (Riley et al., 2008), and Europeana. One immediate revelation was somewhat surprising, however. The greatest difference between collections—and the source of the most difficulties—is not the metadata schemas employed or repositories used, but the extent to which simple metadata, like unqualified Dublin Core exposed over OAI-PMH, must be processed, and, more importantly, how metadata is input and managed locally.

When data is shared in MODS, MARCXML, or even qualified Dublin Core, the richness and completeness of the records transfers relatively easily to the DPLA model. Not surprising, of course, is that the more granular the original record, the better the output at the other end. However, unqualified Dublin Core—most often exposed over OAI-PMH—requires a great deal more analysis and a greater number of complex transformations to identify and map discrete values in a single field to multiple fields in the MAP. For example, specific transformation and enrichment modules are created to determine when a dc:coverage field contains only spatial information, spatial information together with temporal information, or only temporal information. Similar issues, although no less challenging, arise from the varied interpretation of values in dc:source, dc:contributor, dc:relation, dc:type, and others. In evaluating the importance or the efficacy of these transforms, DPLA is reminded that "minimally descriptive metadata … is still minimally descriptive after multiple quality repairs" (Lagoze, et al. 2006). In some ways, this problem is exacerbated further given that Hubs are often aggregators themselves. The degree to which values have been "dumbed down" is not always well documented in terms of how or where this simplification occurred.

It also became immediately clear when a Hub, or its partners, consistently employed and applied (or didn't) controlled vocabularies. While most Hubs follow general guidelines for geographic names (e.g., selecting terms from vocabularies like TGN or LCSH), they are not always applied consistently. Again, this is in part because many Hubs are themselves aggregators of content from hundreds of partners. On DPLA's long-term roadmap for implementation is the work to implement reconciliation of string literals against large controlled vocabularies. Interestingly, in many collections, Hubs' partner names are not taken from controlled vocabularies, or if they are, either this is not indicated in the data or the authorized form of name lacks important contextual information. This has led to a surprising number of errors or unfamiliar values in the data, at least initially. One Hub utilizes the Library of Congress Name Authority File to create their controlled list of partner names. While on the surface this seems like a prudent approach, until the terms are associated with URIs and are augmented with more information, many of the names have very little meaning outside of their local context. For example, not everyone can readily associate the LC Name Authority "J. Y. Joyner Library" with East Carolina University (the parent institution).

## 4. Responses and requests from DPLA Hubs

DPLA personnel have actively worked in partnership with Hubs to identify and openly communicate quality issues in the data that they are sharing. Hubs have been responsive and often eager to make updates and changes to data and even the mappings in their local systems to better align with international practice and the DPLA data model. All agree that this has meant better data quality at both the local and global level. Through this process, Hubs have shared thoughts on ways that ingest could be improved. In some cases, they have begun local development on tools that transform and enrich their data before it reaches DPLA. Some of the requests DPLA has heard align well with its own internal priorities and needs.

### 4.1. Greater control over and feedback during the ingestion process

As mentioned earlier, the community feels strongly that they would benefit from an "ingestion dashboard" that offers a selection of enrichment processes from which Hubs could choose to apply to their data during the ingest process. Because the Hubs know their data best, enabling access to an ingestion dashboard and involving them as early as possible in the initial mapping process would give the Hubs more control over the way their data is exposed via DPLA. Also, it would shed light on what remains a somewhat opaque process for those who are not proficient with the technologies in use. In the interim, DPLA has developed a basic content quality assurance dashboard for internal use and review by Hubs before an initial ingestion reaches the DPLA production data store. The dashboard application is part of the platform API infrastructure, and provides a stripped-down user interface for search and browse of ingested metadata, and the generation of reports on metadata output from the transformation and enrichment pipeline. In addition, integrating tools that provide better visual representations of how metadata is mapped at ingestion and presented in the DPLA portal interface (e.g., Gregory and Williams, 2014) would benefit stakeholders across the DPLA network.

### 4.2. Access to data quality reports

As part of the initial ingestion process for a new Hub, a series of reports are produced that enable DPLA staff to review the values in each field mapped to the DPLA application profile. For each property, two reports are produced: an itemized list of all values in the field and the corresponding DPLA record identifier, and a count of all of the values in that field. The reports are produced from the enriched data, after geocoding and normalization have been applied. Some Hubs, especially those with repository systems that cannot easily generate aggregated reports for a given element or predicate, have requested access to reports on their *unprocessed* data. This would allow them to assess their metadata and perform remediation locally, before it is ever harvested by DPLA. While valuable, this will require significant re-engineering of the ingestion system before it can be implemented.

### 4.3. Upstream data flow: receiving DPLA-provided enrichments

The greatest challenge, but one that several Hubs have voiced interest in investigating, is a method for applying enrichments undertaken by DPLA as part of the ingestion process back to their local data sets. While DPLA provides data dumps for all Hubs' metadata both as individual and collective compressed dump files on the DPLA portal, working with this data can be challenging due—in part—to the sheer size of the files. For Hubs that have a strong technology team and a software environment that would allow it, pulling data from the DPLA API and merging changes with their local data might be a possibility. For others, especially those using systems like CONTENTdm that do not allow for the expression of relationships between fields, this will likely remain an impossibility. Nonetheless, to provide this service in a scalable fashion will require DPLA to better track how and when enrichments are applied, and when they may or may not be necessary.

### 4.4. Further tool and infrastructure development

While DPLA provides guidance to Hubs about particular standards, schemas, or protocols used to standardize, aggregate, and/or provide metadata, DPLA does not usually recommend or require use of any specific tools or applications to harvest, transform, or enrich metadata. Some Hubs have expressed an interest in working with other Hubs or with DPLA to develop tools to help with these processes. Even when formal collaboration has not yet been established, DPLA now finds itself providing an important service, mediating connections across Hubs to identify when the community faces common challenges.

## 5. Planning for needed improvements

Based on this feedback from Hubs, as well as needs identified through the challenges listed previously, DPLA is now reassessing its priorities and planning to address these issues. In some cases, resolving these issues may directly impact the infrastructure DPLA has in place, and addressing others clearly relates the need for DPLA to identify the level to which it should provide services on behalf of its Hubs. Some of the major areas of focused effort over the next year include the following.

### 5.1. Revision of the DPLA Metadata Application Profile

While the Metadata Application Profile is based on the Europeana Data Model (EDM), it has nonetheless diverged from it due to the pressures of DPLA's initial launch outlined above. Accordingly, DPLA is undertaking revision of the MAP to bring it back to closer alignment with EDM, which will allow the ingestion process to better associate URIs with given predicates in the MAP. As indicated in section 2.1, DPLA had sufficient needs that led to the development and implementation of MAP version 3.1. As an organization, DPLA has committed to reviewing the MAP on an ongoing basis, and is already planning for further changes to be included in MAP version 4. These include shifting to the class defined by EDM for spatial data (edm:Place), better support for controlled vocabularies for subject and genre statements, and investigating the addition of a class to provide support for annotation information. Future versions will also allow DPLA and other consumers of the ingested metadata to better incorporate annotations, either in the form of user-generated metadata, or automated output based on the results of transforms and enrichments during each ingest process.

### 5.2. Reassessment of "data quality" and "validation" in the context of DPLA

To provide better tools that ensure the validity and quality of metadata, there will need to be a clear understanding of how those terms are defined in the context of the DPLA/Hub collaboration. Lagoze et al. (2006) suggest that safe transforms are not necessarily scalable, and as such, DPLA and its Hubs must work together to clearly identify which remediation or augmentation processes add the most value to partners and other stakeholders. In addition, DPLA needs to determine whether validation against the MAP is a priority, and to have a clearer delineation of which party must provide the appropriate source data to fulfill the obligations of the MAP (i.e., DPLA, the Hub, or the partner). If explicit validation against the MAP becomes a priority for DPLA and its stakeholders, it will likely require the addition of a means to validate a set of statements against the constraints of the MAP as an RDF application profile. As a preliminary investigation, the co-authors have contributed use cases to the DCMI RDF Application Profiles Task Group.

### 5.3. Encouraging Hubs to undertake metadata transformation and enrichment locally and to develop appropriate tools

Since Hubs often know their metadata (and that of their partners) best, DPLA sees promise in Hubs taking on greater responsibility for metadata remediation, enrichment, and transformation to the MAP at the local level whenever possible. In many cases, DPLA has seen leadership in this area from Service Hubs, in particular (organizations or collaborative endeavors that aggregate metadata and provide services to several cultural heritage organizations, usually at a state or regional level). Some Service Hubs are already actively developing open source software to support these processes. Ultimately, software and infrastructure developed by the Hubs may benefit DPLA and its network further if it can be easily reused.

There are several notable examples of this leadership shown by Service Hubs. Developers at the Boston Public Library (2014) have developed a Ruby module for improved geocoding and reconciliation of geographic names against vocabularies, which is used to augment both their own data as well as data aggregated by Digital Commonwealth, the Service Hub for Massachusetts. University of Minnesota Libraries (2014a, 2014b, 2014c) are developing a suite of tools to harvest, transform, and augment metadata for materials aggregated by the Minnesota Digital Library, with the ultimate goal to provide DPLA with the metadata compliant with the MAP. In addition, the North Carolina Digital Heritage Center (NCDHC) has gained significant expertise in using REPOX for metadata aggregation as a DPLA Service Hub and has developed additional quality assurance applications to support this work (Gregory and Williams, 2014). In addition, to promote reuse, NCDHC released these as open source applications on GitHub. The tools allow NCDHC staff to review mappings, check for the presence of required properties or elements (NCDHC 2014a), and to provide a preview simulating the DPLA's portal user interface for individual new records that can be reviewed by their partners (NCDHC 2014b).

### 5.4. Improvement of documentation for metadata model and ingestion process

Despite both metadata mapping documentation and the code for the ingestion system being publicly available, there is still a significant gap in terms of materials available to understand the DPLA ingestion process. Accordingly, DPLA has begun to address this need by releasing an introductory white paper that explains the MAP (Digital Public Library of America 2014c) and creating a wiki page that collocates existing documentation about metadata, partnerships, and related activities (2014d). DPLA continues to develop further documentation that describes the ingestion process. This work will also likely give DPLA staff better insight about the expectations for these processes. In addition, DPLA staff has also supplemented the MAP version 3.1 documentation with explicit references to how properties within MAP are serialized as JSON-LD.

### 5.5. Improvement or replacement of the DPLA ingestion system

Many of the issues identified by DPLA demonstrate that the current ingestion system, while suitable as a prototype platform for the harvesting, remediation, mapping, and enhancement from many sources, is not entirely suited to the needs of a large-scale aggregator. Internally, DPLA staff has been working to address some issues while investigating whether a substantial refactor or a complete replacement would better serve the needs of the organization. A few areas for immediate focus include increasing efficiency, providing better automation, allowing DPLA content staff to oversee and understand the ingestion process directly with less mediation by the DPLA technology staff by the development of the aforementioned ingestion and QA dashboards, and more clearly defining the shared set of transforms and enrichments for all sources. In addition, the use of domain specific languages that are purpose-built for metadata mapping, transformation, and enhancement holds promise (e.g., Phillips, Tarver and Frakes, 2014 and LibreCat, 2014). These changes, in turn, could allow DPLA to create a system with its Hubs that is more approachable and transparent for those less comfortable with command-line applications and the orchestration of web services. DPLA has not committed to specific candidates for a replacement or undertaken extensive requirements analysis for a new ingestion system. Nonetheless, DPLA is interested in investigating both the previously described software suite under development by University of Minnesota, as well as Supplejack, the harvesting and augmentation framework used by DigitalNZ (2014).

## 6. Conclusion

Despite ongoing challenges with its existing infrastructure, DPLA has successfully aggregated over seven million records from 20 Hubs and nearly 1,300 partner institutions. The lightweight infrastructure used to support ingestion, storage, and indexing allowed the technical implementation team to quickly develop a system to harvest, remediate, and enrich metadata in varying formats. While the current ingestion system clearly has limits, the experience has allowed DPLA and its Hubs to identify shared needs and opportunities for collaboration while adding value to metadata for digitized cultural heritage materials. As the partnership around DPLA grows, the organization is uniquely situated to foster a community of practice that develops and provides documentation, software, and a forum to address ongoing needs in the remediation and enhancement of metadata at a national scale.

## Acknowledgements

The Digital Public Library of America wishes to thank and acknowledge the support of the following organizations that have funded its efforts: The Alfred P. Sloan Foundation; The Andrew W. Mellon Foundation; The Arcadia Fund; The Bill & Melinda Gates Foundation; the Institute of Museum and Library Services; The John S. and James L. Knight Foundation; The Mrs. Giles Whiting Foundation; and the National Endowment for the Humanities.